\documentclass[useAMS,usenatbib]{mn2e}
\usepackage{graphicx}
\usepackage{lscape}
\usepackage{rotating}
\usepackage{colordvi}
\usepackage{color}
\usepackage[english]{babel}
\usepackage{amsmath}
\usepackage{amssymb}
\usepackage{amsfonts}
\usepackage{epsfig}
\usepackage{subfig}
\usepackage{natbib}

\title[Constraining $f(R)$ gravity with PLANCK data on galaxy cluster profiles]
{Constraining f(R) gravity with PLANCK data on galaxy cluster profiles}
\author[I. De Martino, M. De Laurentis, F. Atrio-Barandela, 
S. Capozziello]{I. De Martino$^{1,4}$\thanks{e-mail address: ivan.demartino@usal.es}, 
M. De Laurentis$^{2,3,4}$, F. Atrio-Barandela$^{1}$, S. Capozziello$^{3,4}$\\  
$^{1}${F\'{\i}sica Te\'orica, Universidad de Salamanca, 37008 Salamanca, Spain;
email: ivan.demartino@usal.es; atrio@usal.es}\\
$^{2}$ {Department of Theoretical Physics, Tomsk State Pedagogical University (TSPU), pr. Komsomolsky, 75, Tomsk, 634041, Russia}\\
$^{3}$ {Dipartimento di Fisica, Universit\`a di Napoli "Federico II"}\\
$^{4}$ {INFN sez. di Napoli Compl. Univ. di Monte S. Angelo, Edificio G, 
Via Cinthia, I-80126 - Napoli, Italy} }

\begin{document}
\date{Accepted xxxx Yyyyber zz. Received xxxx Yyyymber zz; 
in original form xxxx Yyyyber zz}
\pubyear{2013}

\maketitle
\label{firstpage}

\begin{abstract}
Models of $f(R)$ gravity that introduce corrections to the Newtonian
potential in the weak field limit are tested at the scale of galaxy clusters.
These models can explain the dynamics of spiral and elliptical galaxies 
without resorting to dark matter. We compute the pressure profiles 
of 579 galaxy clusters assuming that the gas is in hydrostatic 
equilibrium within the potential well of the modified gravitational field.
The predicted profiles are compared with the average profile obtained by stacking
the data of our cluster sample in the Planck foreground clean map SMICA. We find
that the resulting profiles of these systems fit the data 
without requiring a dominant dark matter component, with model parameters 
similar to those required to explain the dynamics of galaxies. Our results
do not rule out that clusters are dynamically dominated by Dark Matter but
support the idea that Extended Theories of Gravity could provide an 
explanation to the dynamics of self-gravitating systems and to the present 
period of accelerated expansion, alternative to the concordance cosmological model.
\end{abstract}

\section{Introduction} \label{One}

Measurements based on Supernovae Type Ia (SNeIa) have indicated that 
the Universe has entered a period of accelerated expansion
\citep{riess, astier, clocc}. Data on Cosmic Microwave
Background (CMB) temperature anisotropies measured by the
Wilkinson Microwave Anisotropy Probe (WMAP)
\citep{wmap9} and the Planck satellite
\citep{PLANCKXV2013, PLANCKXVI2013, PLANCKXX2013, PLANCKXXI2013}, on
Baryon Acoustic Oscillations (BAO) \citep{bao:blake} and other observables,
together with SNeIa data, favor the concordance
$\Lambda$CDM model. In this model, the energy component that drives
the current period of accelerated expansion is a cosmological constant $\Lambda$.
The associated energy density is $\Omega_\Lambda\simeq 0.7$, in units of the critical
density. The second most important component is Dark Matter (DM),
a matter component required to explain the formation of galaxies and 
the emergence of Large Scale Structure, with 
$\Omega_{DM}\simeq 0.26$. More general models assume that the
acceleration is due to an evolving form of  Dark Energy (DE)  characterized
by an equation of state parameter $\omega\le -1/3$. For these
models, cosmological observations indicate that $w=-1.13^{+0.13}_{-0.10}$
\citep{PLANCKXVI2013}, fully compatible with a cosmological constant ($w=-1$).

As an alternative, models involving extensions of General Relativity (GR)
have also been widely considered ({ for comprehensive reviews see 
\citet{carroll}, \citet{sotiriou}, \citet{defelice}, \citet{Nojiri2007}, \citet{Nojiri2011}} and \citet{PRnostro}). 
In this approach, the Hilbert-Einstein action 
changes from being linear in the Ricci curvature scalar, $R$, to a more 
general function. The simplest extensions are $f(R)$ models, where the
Lagrangian is a function (possibly analytic) of the Ricci scalar.
In these models, the higher order gravity terms introduced in the action
are responsible for the present period of accelerated expansion. In some
Extended Theories of Gravity (ETG), the Newtonian limit is 
also modified and models have been constructed where the dynamics of 
galaxies can be explained without requiring a DM component.  
For instance, analytic $f(R)$ models give rise to 
Yukawa-like correction to the gravitational potential 
\citep{PRnostro, annalen} that do not require DM to explain the 
flat rotation curves of spiral galaxies \citep{cardone} or
the velocity dispersion of ellipticals \citep{napolitano}. 
The constraints derived from planetary dynamics are weak since the Yukawa
correction is negligible at those scales \citep{trosi, tartaglia, berry}. 
ETG also modify the hydrostatic equilibrium
of stars: \citet{arturo}, \citet{farinelli} have compared the Lan\'e-Endem solution of
polytropic gases in both $f(R)$ and general relativity and found them to
be compatible while \citet{jeans} analyzed Jeans instabilities in
self-gravitating systems and studied star formation in $f(R)$ gravity.

Since the exact functional form of the Lagrangian is unknown, theoretical
considerations need to be complemented with observations. Thus, it is important 
to test potential models using all available data. At present, clusters of 
galaxies are the largest virialized objects in the Universe and offer the 
opportunity to test these alternative theories of gravity on scales larger 
than galaxy scales. Using the mass profiles of clusters of galaxies, \citet{salzano} 
showed that ETG provide a fit to the distribution of baryonic matter (stars+gas) 
derived from X-ray observations in 12 clusters without requiring DM.
Nevertheless, in conventional cosmological models, the non-linear evolution and 
virialization of self-gravitating objects is studied using numerical 
simulations. $f(R)$ models have a much larger number of degrees 
of freedom and the study of galaxy and cluster formation requires
more complex  simulations, specific for each particular Lagrangian. 
A first attempt to constrain ETG using cluster abundances in 
numerical simulations has been carried out by \citet{hu1} and \citet{hu2}.
{ 
Other numerical constraints on $f(R)$ models can be found in
\citet{song}, \citet{sawicki}, \citet{hu2007-1,hu2007-2}
and \citet{lima}.}
More promising is the study of temperature fluctuations on the CMB.
Galaxy clusters are reservoirs of hot gas that induces
anisotropies by means of the Sunyaev-Zeldovich (SZ) effect \citep{tsz,ksz}. 
Pressure profiles of galaxies can be computed in ETG assuming that
the gas is in hydrostatic equilibrium within the potential well of clusters. 
This is in agreement with the results of 
numerical simulations based on the concordance cosmology that showed 
that gas is in hydrostatic equilibrium in the intermediate regions of clusters,
while in the cluster cores, the physics of
baryons is more complex and in the outer regions it is dominated 
by non-equilibrium processes \citep{kravtsov12}. 
{ Recently, hydro-numerical simulations are being carried out to
study the properties of galaxy clusters and groups in ETG. 
\citet{arnold} showed that the intracluster medium temperature increases
in $f(R)$ gravity in low mass halos but the difference disappears in massive
objects. Based on these results we will assume that the physics of the gas
will be weakly dependent on the underline theory of gravity.} 

The SZ anisotropies generated by individual clusters and 
by the unresolved cluster population have been measured by
the Atacama Cosmology Telescope (ACT) \citep{ACT1,ACT2,ACT3,ACT4}, 
the South Pole Telescope (SPT) \citep{SPT1,SPT2,SPT3,SPT4}
and the Planck satellite \citep{PLANCKV2012, PLANCKX2012, PLANCKXX2013, 
PLANCKXXIX2013}.  Gas profiles based on the Navarro-Frenk-White 
(hereafter NFW, \citet{NFW1997}) profile, derived from numerical simulations, 
have been found to be in agreement with TSZ
\citep{atrio2008} and X-ray observations \citep{arnaud2010}.
Nevertheless, the contribution of the unresolved cluster
population in WMAP 7yr data has been found to be smaller than expected 
based on theoretical and numerical modeling of clusters \citep{wmap7}.
For the Coma cluster, the analysis of Planck data \citep{PLANCKV2012, PLANCKX2012}
finds a normalization of $\sim  10-15\%$ lower
compared with the parameters derived from XMM observations. 
These discrepancies can be related to the existence of complex
structures and substructures in clusters of galaxies as well as
to the limitations of the theoretical modeling \citep{fusco2012},
that is the approach we are going to consider here.

In this article, we will compare the pressure profiles of clusters of
galaxies in $f(R)$ models with Planck data. To construct the pressure
profiles, we will assume that the gas is in hydrostatic equilibrium
within the potential well generated by the cluster. At this level,
our assumption can not be applied to models not in equilibrium 
like the Bullet cluster \citep{clowe}. We will restrict our analysis
to $f(R)$ models of gravity that introduce Yukawa corrections 
to the Newtonian potential in order to test if the dynamics
of clusters of galaxies can be also described without a 
dominant dark matter component.
The paper is organized as follows: in Sec.~2, we consider the weak 
field limit of $f(R)$ gravity deriving the gravitational potential for 
self-gravitating objects; in Sec.~3, we present the 
pressure profiles based on the NFW profile and X-ray data most
commonly used and we compute the pressure profile for $f(R)$ models; in Sec.~4 
we describe the data used in our analysis; 
in Sec.~5, we discuss our results and, finally, in Sec.~6 we present
our main conclusions.

\section{Yukawa corrections to the Newtonian potential
in $f(R)$-gravity} \label{sect:2}

In $f(R)$ ETG, field equations are derived from the action 
\begin{equation}\label{action}
{\cal A}=\frac{c^4}{16\pi G}\int d^4x \sqrt{-g}f(R) + {\cal L}_{m}\, ,
\end{equation}
yielding
\begin{equation*}\label{eq:fe1}
f'(R)R_{\mu\nu}-\frac{f(R)}{2}\,g_{\mu\nu}-
f'(R)_{;\mu\nu}+g_{\mu\nu}\Box_g f'(R)\,=\,8\pi G T_{\mu\nu}\,,
\end{equation*}
where 
${\displaystyle f'(R)={df(R)}/{dR}}$ is the first derivative with 
respect to the Ricci scalar, $\Box_g={{}_{;\sigma}}^{;\sigma}$ 
is the d'Alembertian with covariant derivatives, 
$\displaystyle T_{\mu\nu}=-2(-g)^{-1/2}{\delta(\sqrt{-g}\mathcal{L}_m)/\delta g^{\mu\nu}}$ 
is the matter energy-momentum tensor, $T$ its trace, $g$ the determinant of
the metric tensor $g_{\mu\nu}$. Greek indices run from $0$ to $3$. 

We search for spherically symmetric solutions of the form
\begin{equation}
 ds^2=g_{tt}c^2dt^2 - g_{rr}dr^2 - r^2d\Omega,
\end{equation}
where $d\Omega$ is the solid angle. Let us restrict our study
to those $f(R)$-Lagrangians that 
can be expanded in Taylor series around a fixed point $R_0$ 
\begin{equation}\label{eq:sertay}
f(R)=\sum_{n}\frac{f^n(R_0)}{n!}(R-R_0)^n\simeq
f_0+f'_0R+\frac{f''_0}{2}R^2+...\,.
\end{equation}
The fixed  point represents the Ricci-scalar in GR for the same
mass distribution. In this case
$f_0$ is a cosmological constant and $f'_0=1$.
Then, the field Eqs.~(\ref{eq:fe1}) can be solved at different orders
in terms of the Taylor expansion. In the Newtonian limit the first
correction  is of order $c^2$. The metric tensor can be written as 
\begin{align}
\label{eq:gtt} g_{tt}& \simeq 1+ 2\Phi_{grav}(r),\\
\label{eq:grr} g_{rr}& \simeq -(1+\Phi_{N}(r)),\\
\label{eq:gthth} g_{\theta\theta}& \simeq -r^2,\\
\label{eq:phph} g_{\phi\phi}& \simeq -r^2\sin^2\theta,
\end{align}
where
\begin{equation}
\Phi_N(r)=-\frac{GM(r)}{r},\;
\Phi_{grav}(r)=\frac{\Phi_N(r)}{(1+\delta)}\left(1+\delta e^{-\frac{r}{L}}\right),
\label{eq:pot}
\end{equation}
Analytic $f(R)$ models that modify the Newtonian limit can be seen
as alternative to Dark Matter. The Yukawa correction to the gravitational
field allows us to study the dynamics of galaxies without requiring dark
matter. The parameters ($\delta$, L) are related to the coefficients
in the Taylor expansion as: $f'_0=1+\delta$ and $L=[-{f'_0}/{(6f''_0)}]^{1/2}$,
where $\delta$ represents the deviation from GR at zero order and $L$ the 
scale length of the self-gravitating object \citep{PRnostro, annalen}.
In the limit $\delta=0$, we recover the Newtonian limit of GR, irrespective
of the scale parameter $L$. In ETG, $L$ depends on 
the scale of the system considered; it  assumes  different values  
for the various self-gravitating systems like 
galaxies or cluster of galaxies while its effects are totally 
negligible at Solar System scales where GR, i.e. the Newtonian limit, 
is totally restored \citep{annalen}. 

The physical meaning of the characteristic length $L$ deserves further discussion. 
As pointed out in \citet{PRnostro}, $L$ can be seen as an extra {\it gravitational 
radius} similar to the Schwarzschild radius. Compared with
GR, that is a second order theory, $f(R)$ gravity is fourth-order and
contains a larger number of degrees of freedom that,
in the weak field  limit, give rise to a new characteristic scale length. 
The paradigm can be extended to $(2k+2)$-order theories of gravity so
any further two derivation orders imply a new characteristic length in 
the Newtonian limit (see \citet{hans} for details), resulting in some
important implications for the theory.
First gravity is no longer a scale invariant interaction but 
depends on the size of the self-gravitating systems. In other words, gravitational 
corrections emerge depending on scales.  Second, the Gauss theorem does not hold 
at finite scales but only  asymptotically. This is not a problem since 
Bianchi identities hold for $f(R)$ as for any ETG theory and conservation laws 
are fulfilled like in GR. 
Third, GR is totally restored at Solar System scales so $f(R)$ theory agrees with 
standard classical tests \citep{tsuji}. Finally, the approach allows to represent 
DM effects only by gravity without requiring new ingredients at the fundamental level. 
This fact could be considered as an astrophysical testbed for relativistic theories 
of gravity since the additional gravitational length $L$ introduced in this model
could be accurately matched with observational data as we are going to show below.

\section{Cluster pressure profiles in $f(R)$ gravity}\label{sect:3}

\begin{center}
\begin{table}
\begin{tabular}{|lccccc|}
\hline
\emph{Model} & $c_{500}$ & $\alpha_a$ & $\beta_a$ & $\gamma_a$ & $P_0$ \\ 
\hline
\emph{Arnaud} & 1.177 & 1.051 & 5.4905 &  0.3081 & $8.403h_{70}^{3/2}$ \\ 
\emph{Planck} & 1.81 & 1.33 & 4.13 & 0.31 & 6.41 \\
\emph{Sayers} & 1.18 & 0.86 & 3.67 & 0.67  &  4.29 \\
\hline
\hline
\emph{$\beta$} & $\beta$ & $n_{c,0}/m^{-3}$ & $r_c$/Mpc & $T_e$/keV    & \\
\hline
                     & 2/3     &   3860.   & 0.25  &  6.48         & \\
\hline
\hline
\emph{$f(R)$} & $\delta$ & L/Mpc & $\gamma$ & &  \\
\hline
                    &   -0.98  &  0.1    &  1.2     & &  \\
\hline
\end{tabular}
\caption{Parameters of the Generalized NFW, $\beta$ and $f(R)$
models represented in Fig.~\ref{Fig1}; the Generalized NFW pressure
profile data is from \citet{arnaud2010}, \citet{PLANCKV2012} and \citet{Sayers2012},
the $\beta$-model data corresponds to the Coma cluster and
the $f(R)$ profile data is the best fit model to Planck data (see Sec.~5).
}
\label{table1}
\end{table}
\end{center}

When CMB photons cross the potential wells of clusters of galaxies, they are 
scattered off by the electrons of the Intra-Cluster medium, inducing 
secondary temperature anisotropies on the CMB of two different type
by means of the SZ effect: a
thermal  contribution due to the motion of the electrons within the cluster
potential well and kinematic one (KSZ) due to the motion of the cluster
as a whole. The TSZ is the only SZ anisotropy that has been measured
for individual clusters and is given by \citep{tsz}
\begin{equation}
\frac{\Delta T}{T_0}=g(\nu)\frac{k_B\sigma_T}{m_ec^2}\int n_eT_e dl,
\label{eq:sz}
\end{equation}
where $T_e$ is the electron temperature, $n_e$ the electron density and the 
integration is carried out along the line of sight $l$. In Eq.~(\ref{eq:sz}) 
$k_B$ is the Boltzmann constant, $m_ec^2$ the electron annihilation 
temperature, $c$ the speed of light, $\nu$ the frequency 
of observation, $\sigma_T$ Thomson cross section and
$T_0$ the mean temperature of the CMB. Finally, $g(\nu)=x\coth(x/2)-4$
is the frequency dependence of the TSZ effect, with $x=h\nu/KT_0$.

To compute the TSZ anisotropy we need to specify the pressure profile
$n_eT_e$ of clusters. Using X-ray data and numerical simulations, several 
cluster profiles have appeared in the literature: 

$\bullet$ The X-ray emitting region of clusters of galaxies
is well fit by the isothermal $\beta$-model profiles
\citep{cavaliere1976, cavaliere1978}. In this model,
the electron density is given by: $n_e(r)=n_{e,0}[1+(r/r_c)^2]^{-3\beta/2}$,
where the core radius $r_c$, the central electron density $n_{e,0}$, the
electron temperature 
and the slope $\beta$ need to be determined from observations.
From the X-ray surface brightness of clusters, $\beta=0.6-0.8$ \citep{jones1984}. 
In Table~\ref{table1} we give the value of the $\beta$ model parameters
of the Coma cluster.

$\bullet$ Outside the central cluster regions, the $\beta$ model overpredicts
the TSZ contribution \citep{atrio2008}. If the electrons are in hydrostatic 
equilibrium within the potential well of dark matter halos, the pressure profile 
is well describe by a Komatsu-Seljak model \citep{KS2002,atrio2008}. 
More recently, \citet{arnaud2010} proposed a phenomenological parametrization
of the electron pressure profile based on generalized Navarro-Frenk-White 
(GNFW) profiles derived from the numerical simulations of \citet{nagai2007}. 
This profile has the following functional form
\begin{equation}
p(x) \equiv \frac{P_0}{(c_{500}x)^{\gamma_a} 
[1+(c_{500}x)^{\alpha_a}]^{(\beta_a-\gamma_a)/\alpha_a}},
\label{eq:universal_profile}
\end{equation}
In this expression $x$ is the radial distance in units of $r_{500}$,
the radius where the average density is 500 times the critical density,
and $c_{500}$ is the concentration parameter at $r_{500}$. Different
groups have fit the model parameters $[c_{500},\alpha_a, \beta_a, \gamma_a, P_0]$ 
to X-ray or CMB data; their best fit values are given in Table~\ref{table1}.

\begin{figure}
\centering 
\includegraphics[width=0.95\columnwidth]{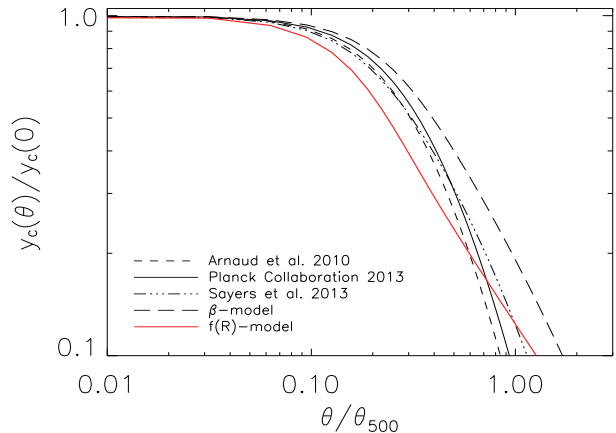}\\
\caption{
Pressure profiles integrated along the line of sight
for the Coma cluster. We represent three GNFW profiles
(dashed, solid and dash-dotted lines), one $\beta=2/3$ model (long
dashed line) and a $f(R)$ model (red solid line). The model parameters
are given in Table~\ref{table1}. The angular diameter distance
is that of the Coma cluster ($z=0.023$).
}
\label{Fig1}
\end{figure}

{ GNFW models fit the DM distribution in numerical simulations
that use newtonian gravity and therefore can not be used to describe
the dynamics in the ETG we are considering. 
Instead,} baryons reside in the potential well
of clusters. The Yukawa correction to the Newtonian potential of 
eq.~(\ref{eq:pot}) modifies the gravitational structure of clusters 
and there is not longer any need to introduce dark matter to explain 
their dynamics. In this limit, to compute
the pressure profile $n_eT_e$ of Eq.~(\ref{eq:sz}) we assume  that
the gas is in hydrostatic equilibrium within the 
(modified) potential well of the cluster
       \begin{equation}\label{eq:HE}
        \frac{dP(r)}{dr}=-\rho(r)\frac{d\Phi_{grav}(r)}{dr},
       \end{equation}
and to describe the physical state of the gas
we further assume that it follows a polytropic equation of state 
       \begin{equation}\label{eq:PES}
        P(r)\propto\rho^\gamma(r).
       \end{equation}
Eqs.~\eqref{eq:HE} and \eqref{eq:PES} together with mass conservation
        \begin{equation}\label{eq:EMC}
        \frac{dM(r)}{dr}= 4\pi\rho(r),
       \end{equation}
and the cluster gravitational potential given by eq.~(\ref{eq:pot})
form a close system of equations that can be solved 
numerically to obtain the pressure profiles of any given cluster as a function
of two gravitational parameter $(\delta, L)$ and the polytropic index $\gamma$.
For illustration,
in Fig.~\ref{Fig1} we plot the different profiles integrated along the line
of sight with the parameters given in Table~\ref{table1}. 
We particularize the models for the Coma cluster.
For convenience, all distances are written in units of $r_{500}$
and the angular scale is $\theta_{500}=r_{500}/d_A^{Coma}$
where $d_A^{Coma}$ is the angular diameter distance of Coma.
Dashed, solid and dash-dotted lines correspond to GNFW profiles with
the \citet{arnaud2010}, \citet{PLANCKV2012} and \citet{Sayers2012} parameters, 
respectively. The long-dashed line corresponds to the
$\beta$ model and the red solid line to the $f(R)$ model.

\section{Data.}

To constrain the ETG model described in Sec.~2,
we will use the pressure profiles of clusters of galaxies
given in Sec.~3. To that purpose we shall use Planck data and
a proprietary cluster catalog.

\subsection{The Cluster Catalog}

Our cluster catalogue contains 579 clusters selected
from ROSAT All Sky-Survey (RASS). Those clusters are outside the minimal Planck mask
that removes a $\sim 20$\% of the sky in the Plane of the Galaxy. 
Clusters are drawn from the three flux limited cluster samples: the
extended Brightest Cluster Sample
(eBCS, \citep{ebeling98, ebeling00}), the ROSAT-ESO Flux Limited X-ray 
catalog (REFLEX, \citep{bohringer1}), and the
Clusters in the Zone of Avoidance (CIZA, \citep{ebeling02,kocevski07}).
For each cluster, the catalog lists  position, flux, and luminosity 
measured directly from RASS data and spectroscopically
measured redshifts. The X-ray electron temperature is derived from 
the $L_X-T_X$ relation of \citet{white}. For each cluster, the
spatial profile of the X-ray emitting gas is fit to a $\beta$-model
convolved  with the RASS point-spread function to the RASS data.
Due to the poor sampling of the surface brightness profile for all
but the most nearby clusters, $\beta$ is fixed to the canonical value of $\beta=2/3$
\citep{jones1984} but the core radii $r_c$ and central electron densities
$n_{e,0}$ are derived from the data. Thus, our catalog provides enough information
to compute the Comptonization parameter of the X-ray emitting region
of all the clusters in our sample. In \citet{atrio2008}, it was found that
the predicted values and those measured in WMAP 3yr data
were in agreement with the $\beta$ model for the inner
part of the clusters, being the discrepancy between the TSZ prediction
and observation below 10\%.

\subsection{Cosmic Microwave Background data.}

\begin{figure}
\centering 
\includegraphics[width=0.95\columnwidth]{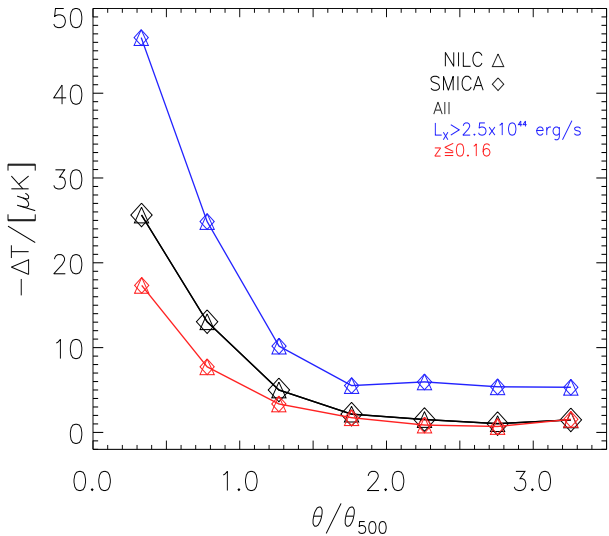}\\
\caption{Average temperature anisotropy 
in the SMICA (diamonds) and NILC (triangles) maps at the position
of two cluster subset selected according to luminosity and redshift
(blue and red lines) and of our full sample (solid black line).
}\label{Fig2}
\end{figure}

The release of WMAP 9yr data \citep{bennett} at the end of 2012 was
followed by the first data release of the Planck satellite in April
2013. Nine maps spanning a frequency range from 32 to 845GHz have
been made publicly available by the Planck 
Collaboration\footnote{{http://irsa.ipac.caltech.edu/Missions/planck.html}}.
While the WMAP team provided foreground clean maps of all Differencing
Assemblies (DA), the Planck Collaboration did not validate foreground
clean maps at all frequencies. Instead, they used component
separation methods to construct a map of CMB temperature anisotropies
combining the data at all frequencies \citep{PLANCKXII2013}. 
The SMICA map was produced by combining all nine Planck frequency maps, 
previously upgraded to the same resolution of $5'$, in spherical 
harmonic space using different weights at different multipoles. 
The NILC map was constructed in needlet space given different
weights to the multipoles and to the spatial positions of the data
in the sky. These two maps were constructed using different
algorithms and, therefore, it is likely that they will 
differ in amplitude, distribution and spatial properties 
of the foreground residuals \citep{PLANCKXIII2013}. We will
perform our analysis in both foreground clean maps,
both with Healpix resolution $N_{side}=2048$ \citep{healpix},
to test for systematics. 

To compute the TSZ profile of the clusters in our sample,
we average the temperature anisotropy at the cluster positions.
At the cluster center the average is over a disc of radius $\theta_{500}/2$
where $\theta_{500}$ is the angular scale subtended by the $r_{500}$
radius of the cluster. Outside the inner disc, we take the
average on rings of width $\theta_{500}/2$.
The measured value is the averaged over all clusters in our sample.
The angular position $\theta$ we associate to each  
data point is the mean of the angular distance to the center of the cluster
of every pixel in a disc or ring. The root mean square
dispersion around the mean is about $0.1\theta_{500}$ for the central
disc that contains the smallest number of pixels and is $0.05\theta_{500}$ or
smaller for the rings.
In Fig.~\ref{Fig2}, we present the results for the SMICA (diamonds)
and NILC (triangles) maps. We compare the results on both maps
for the full sample (solid black line) and for two cluster
subsets, selected according to luminosity (blue) and
redshift (red). The results of both maps differ by less than
1\% in the three samples proving that the differences in the component 
separation method do not distort the TSZ anisotropy associated with 
clusters. The agreement between the TSZ profiles measured in 
the SMICA and NILC maps demonstrates that systematic  
effects will not affect our final results. 

To each data point we associate an error bar obtained
by evaluating 1,000 times the average profiles at 579 random
positions in the SMICA and NILC maps. To avoid overlapping real
and simulated clusters, we excise a disc of 80\arcmin\ around
each cluster in our sample. The errors on both maps
are also indistinguishable. For comparison, we analyzed
the W-band of WMAP 9yr data. The results were very similar
to those of Planck except for larger error bars. 
As remarked in \citet{PLANCKXII2013}, at high latitudes, outside the Galactic
Plane, the amplitude of the foregrounds residuals present on 
the SMICA map is a few $\mu$K, smaller than those
on the NILC map.  Therefore, since NILC or WMAP do not provide extra 
information and since they are more affected by noise or foregrounds
than SMICA, we will restrict our analysis to the latter data.

\subsection{The average SZ profile}

\begin{figure*}
\centering 
\includegraphics[width=2\columnwidth]{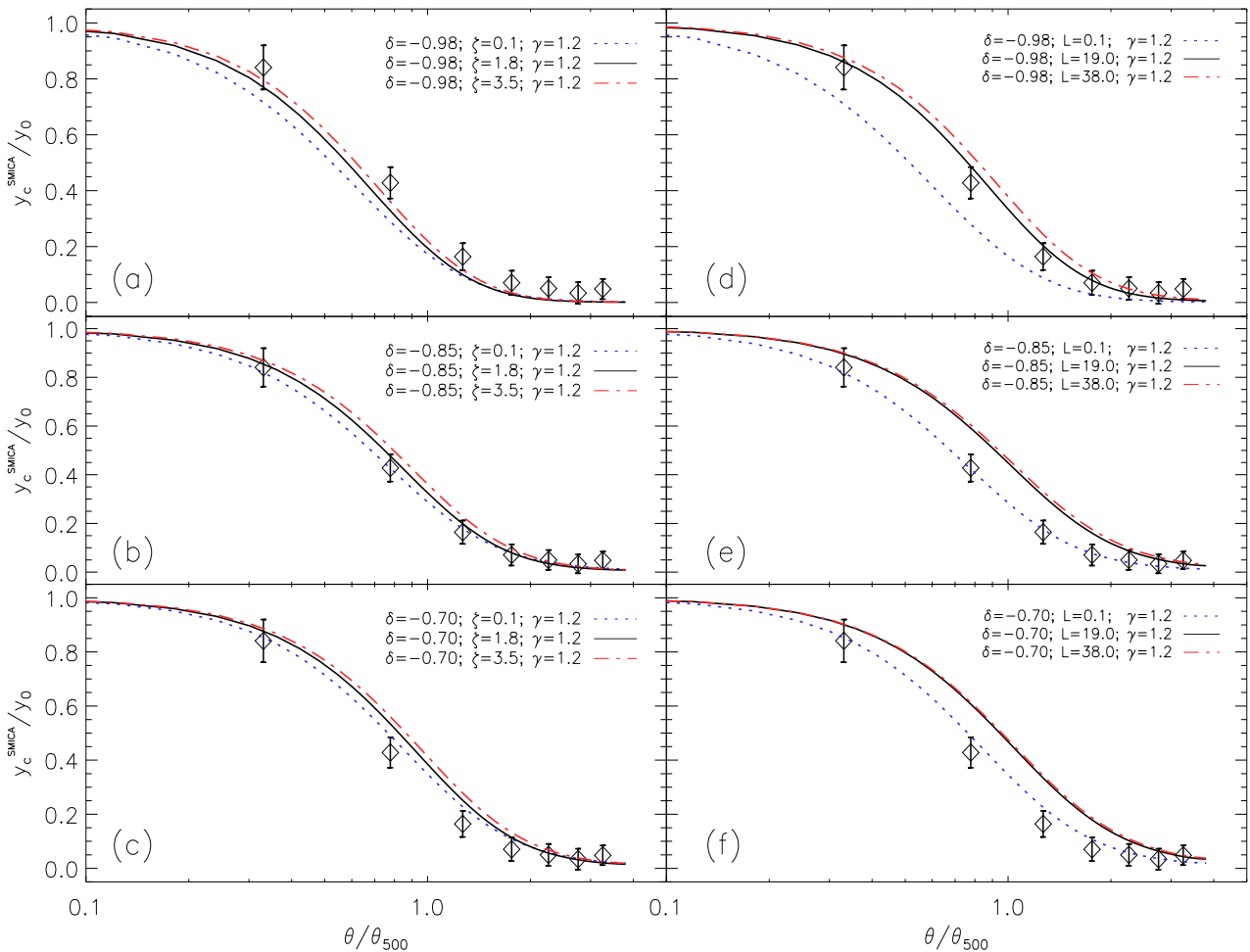}\\
\caption{Pressure profiles of clusters in $f(R)$ gravity and SMICA data.
Panels a-c correspond to the parametrization $L= \zeta r_{500}$ (Model A),
while d-f correspond to the same scale $L$ for all clusters (Model B).
}\label{Fig3}
\end{figure*}

To compare cluster profiles with observations, we  measure
the angle subtended by every cluster in units of $\theta_{500}$.
For each cluster, the radial scale $r_{500}$ can be derived using the
following scaling relation \citet{bohringer2}
\begin{equation}
r_{500}=\frac{0.753h^{-1}\rm{\ Mpc}}{h(z)}
\times\left(\frac{L_X}{10^{44}h^{-2}\rm{\ erg\ s}^{-1}}\right)^{0.228},
\label{r500}
\end{equation}
{ The radius $r_{500}$ will allow us to test if the characteristic scale
of our ETG, $L$, depends on the cluster properties or not. We checked that
our results did not depend on the uncertainties of eq.~\eqref{r500} and we
will not consider them any further. Similarly,
we did not consider other scaling relations based on different data 
\citep{piffaretti,PLANCK-ERX}}. Eqs.~\eqref{eq:pot}, \eqref{eq:HE}, 
\eqref{eq:PES} and \eqref{eq:EMC} allow us to compute the pressure
profile of all clusters in the data as a function
of three parameters: $(\delta, L,\gamma)$. These profiles are integrated
along the line of sight to be compared with those measured in the SMICA map. 
As indicated in Sec.~2, $L$ characterizes the
dependence of $f(R)$ gravity on the size of the gravitating system.
We consider two parameterizations of $L$ to test if the theory
depends on the properties of the clusters:
(A) $L= \zeta r_{500}$ is different for each cluster but depends
homogeneously on $r_{500}$ for the whole sample and
(B) where $L$ is the same for all clusters.
In Fig.~\ref{Fig3} we plot the pressure profile 
integrated along of line of sight, convolved with a gaussian
beam of 5\arcmin\ resolution, for different model parameters. 
Our models only predict the profile but not the central
anisotropy. For this reason, we normalize all our theoretical
profiles to unity. The data is equally normalized by dividing
all the averages by the mean temperature on a disc of $0.1\theta_{500}$
radius. Error bars are computed in the same manner, renormalizing the
disc and rings at random positions on the sky by the mean
on the central disc of $0.1\theta_{500}$.
In Figs.~\ref{Fig3}a-c $L$ is different 
for each cluster (Model A) and in Figs.~\ref{Fig3}d-f $L$
is the same for all clusters (Model B). To avoid overcrowding
the plots, we fixed $\gamma=1.2$. In each panel we
show the variation of the pressure profile with $L$. 
Notice that in Model B, when $L\ge 20$Mpc, the variations on the 
profile are small. This is logical
since $L$ is the scale length of the Yukawa correction, that becomes
negligible for large values of $L$. For illustrative purposes
we overplot the SMICA data shown in Fig.~\ref{Fig2}, normalized to unity,
with their corresponding error bars. 

\section{Results and discussion}\label{sect:5}

To determine the model parameters that best fit the SMICA data
we generate pressure profiles for different values of the
parameters $(\delta,L, \gamma)$, integrated along the
line of sight and convolved with a Gaussian beam
with the same resolution of the SMICA map to compare
them with the data. On physical grounds, we fix
our parameter space to be $\delta=[-0.99,1.0]$ since
if $\delta<-1$ the potential is repulsive and diverges at $\delta=-1$. 
In the parametrization $L= \zeta r_{500}$ we take $\zeta=[0.1,4]$.
When $L$ is the same for all clusters, we fix the interval to be
$L=[0.1,20]$Mpc, from the scale of cluster
core radius to the typical mean cluster separation scale.
Finally, the polytropic equation of state parameter is set to vary
within the range $\gamma=[1.0,1.6]$, that corresponds to an isothermal 
and adiabatic monoatomic gas, respectively. We take 30 equally spaced steps in
all intervals.

\begin{figure*}
\centering 
\includegraphics[width=2\columnwidth]{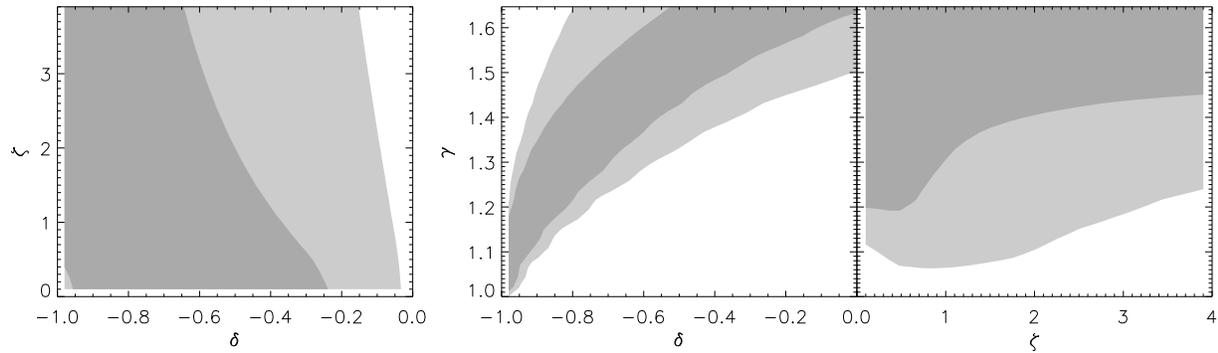}\\
\caption{Confidence contours for pairs of parameters of
Model A. Contours are at the 68\% and 95\% confidence level. 
}\label{Fig4}
\end{figure*}

\begin{figure*}
\centering 
\includegraphics[width=2\columnwidth]{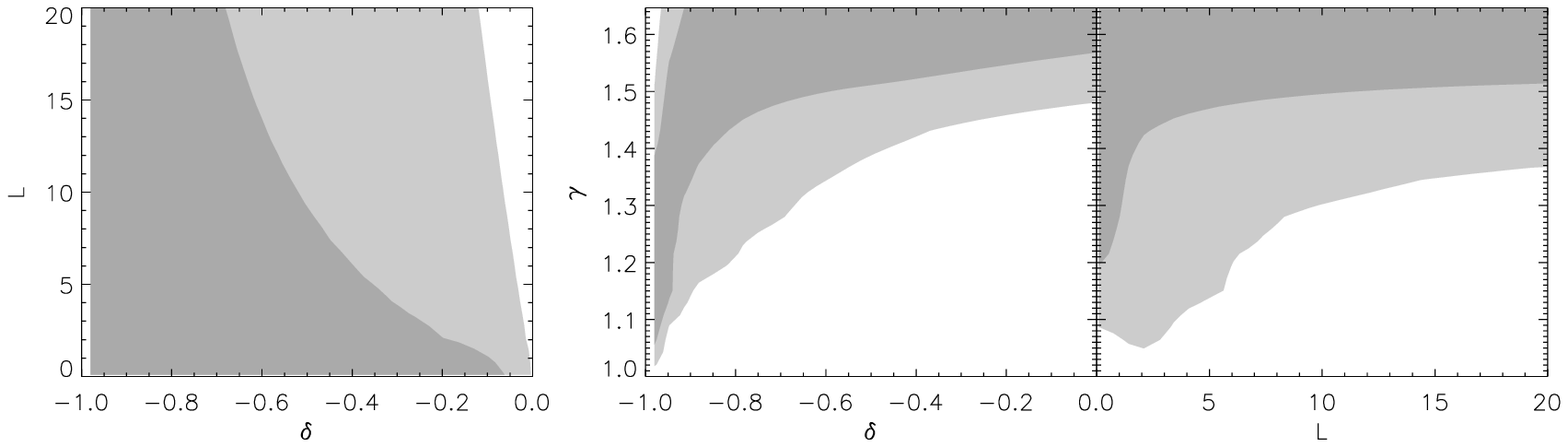}\\
\caption{Same as in  Fig. \ref{Fig4} for Model B.} 
\label{Fig5}
\end{figure*}

In Figs.~\ref{Fig4} and \ref{Fig5}, we compute the confidence contours
for the different model parameters of Model A and Model B, respectively. 
The likelihood function $\log{\cal L}=-\chi^2/2$ is computed as
\begin{equation}
\chi^2 ({ p})=\Sigma_{i,j=0}^{N} 
(y({ p}, x_i)-d(x_i))C_{ij}^{-1}
(y({ p}, x_j)-d(x_j))
\label{eq:chi}
\end{equation}
where $N=7$ is the number of data points. The mean profile $y({ p},x_i)$ 
of all the clusters in our sample depends on three parameters:
${ p}=(\delta,L,\gamma)$. In eq.~(\ref{eq:chi}), $d(x_i)$ is the SMICA average
profile and $C_{i,j}$ is the correlation function between bins.
{ To compute the correlation function we choose 579 random 
positions outside the locations of known clusters and compute the average
temperature anisotropy on discs and rings of size $\theta_{500}$,
different for each of the random clusters. The process is repeated
1,000 times and $C_{ij}$ is the average correlation between bins 
of any given cluster, averaged over all clusters and all simulations.}

{ The value of the model A and B parameters that maximize the likelihood are given
in Table~\ref{table2}. }
In Fig.~\ref{Fig4} we plot the 68\% and 95\% confidence contours for pairs 
of parameters of Model A. Fig.~\ref{Fig5} shows the same contours for the Model B.  
{ Since the models are very similar to each other, the likelihood function
is flat close to the maximum. The $1\sigma$ contours are cut by our physical
boundaries on $\delta$ and $\gamma$. Consequently, 2D contours
of the marginalized likelihoods of pairs of parameters of these Figs are
not closed, and only lower or upper limits to the parameters can be derived
from their marginalize 1D likelihoods. At the 68\% and 95\% confidence
levels those limits are 
$\delta<-0.46,-0.10$, $\zeta < 2.5,3.7$ and $\gamma>1.35,1.12$ for the Model A
parameters and 
$\delta<-0.43,-0.08$, $L<12,19$ Mpc and $\gamma>1.45,1.2$ for the Model B.
In general, model parameters are weakly constrained. In particular, the polytropic 
index constraints dominated by the physical boundary on this parameter.
The characteristic scale length $L$ is similar in both models, whether it scales
with $r_{500}$ or is identical for all clusters. In retrospect, this explains
why the results of model A did not depend on the uncertainties in the scaling 
relation of $r_{500}$, given in  eq.~\eqref{r500}. But even if the parameters
are weakly constrained, let us remark that 
in both models, A and B, the value $\delta=0$ is excluded at 
more than a 95\% confidence level.}
Since $\delta\simeq 0$ 
corresponds to the standard Newtonian potential without DM then the data 
does rule out that baryons alone are the dominant matter
component in clusters. 

The open contours in Figs.~\ref{Fig4} and \ref{Fig5} reflect the
physical limitations of our model.  We can not extend our 
parameter space beyond $\delta=-1$. The limitation stems from the 
use of first order perturbations with respect to a background model. 
The contours show that at the $1\sigma$ level $L$ is compatible with
zero. Physically, at $L\simeq 0$ the gravitational field corresponds
to the Newtonian potential generated by a mass $M'=M/(1+\delta)$.
As $\delta+1\simeq 0$ then $M'\gg M$ and the gravitational field is that of
a system that contains a large fraction of DM distributed like the
baryonic gas. Briefly, while our results show that cluster TSZ profiles
in ETG are compatible with the data, they do not rule out that clusters
could contain a significant fraction of DM. In summary, 
in order to fit the TSZ data, clusters are either dominated by 
DM or the Newtonian potential includes a Yukawa correction.

Comparison of Figs.~\ref{Fig4} and \ref{Fig5} also shows that the data does 
not have enough statistical power to discriminate between Models A and B. 
Importantly, the results are consistent with those obtained by 
\citet{sanders} and \citet{napolitano} using spiral and elliptical
galaxies, respectively. In model A
we find the same correlation between the gravity parameters
$L$ and $\delta$ that in the case of galaxies: to accommodate the data,
larger values of $L$ require lower values of $\delta$, while the
behavior is the opposite in Model B. This different scaling
suggests that Model A is in better agreement with the dynamics
of galaxies than Model B. Also, conceptually is the preferable model
since $L$ scales with the size of the self-gravitating system.
The agreement of the central values
of $\delta$ and $L$ with those of galaxies, that correspond to
a different linear scale, is very reassuring; the dynamics 
of galaxies and clusters can be equally described by ETG, without 
requiring DM. In other words, DM and alternative gravity models are 
equivalent descriptions that could be discriminated only by some 
signature at fundamental scales, i.e. the discovery of new particles 
non-interacting at electromagnetic level, or the clear evidence of 
some new gravitational mode not related to GR (\citet{PRnostro}, \citet{bogdanos}). 

For comparison, we also compute the likelihood of each of
the models given in Table~\ref{table1} and their $\chi^2$ per degree of freedom
are given in  Table~\ref{table2}. For the $\beta$ model we generate
the profile of each cluster using the data of our catalog. The
$\beta$ model does not produce a good fit to the data, in agreement
with our previous results using WMAP data \citep{atrio2008},
since this model only fits the X-ray emitting regions of the
inner parts of clusters.  Comparing
the three GNFW parameters, the \citet{arnaud2010} parameters, derived using 
the X-ray data of 33 clusters, performs better that either
the \citet{PLANCKV2012} or the \citet{Sayers2012} parameters, that were
obtained from TSZ observations.  These discrepancies are not relevant since
we did not explore the parameter space to find the best fit values
of GNFW models to the SMICA data.  Nevertheless, the fact that our
$f(R)$ profiles fit significantly better than any other model
is a clear indication that our assumption of a polytropic gas in hydrostatic
equilibrium in the cluster potential well is supported by the data.

\begin{center}
\begin{table*}
\begin{tabular}{|l|cccccc|}
\hline
 Model     & $\delta$ & L & $\gamma$ & $f'_0$ & $f''_0$ &$\chi^2_{dof}$\\
           &          & (Mpc) &      &        & (kpc$^{-2}$) & \\  
\hline
\hline
Model A ($L= \zeta<r_{500}>$)   & -0.98 & 1.12 & 1.07 & 0.02 & -0.003  & 0.25\\
Model B                          & -0.98 & 1.91 & 1.07 & 0.02 & -0.001 & 0.25\\
\hline
\citet{arnaud2010}              & - & - & - & - & - & 1.38\\
\citet{PLANCKV2012}             & - & - & - & - & - & 2.27\\
\citet{Sayers2012}              & - & - & - & - & - & 7.70\\
$\beta (=2/3)$-model            & - & - & - & - & - & 15.17\\
\hline
\end{tabular}
\caption{$\chi^2$ per degree of freedom 
($\chi^2_{dof}$) for the $\beta$-model, GNFW models with parameters
given in Table~\ref{table1}, and for the two $f(R)$-parametrization considered 
in this work.}\label{table2}
\end{table*}
\end{center}

\section{Conclusions}\label{sect:6}

We have constructed cluster pressure profiles based on the Yukawa-like 
correction to the Newtonian potential obtained in the weak field 
approximation of $f(R)$ gravity. These models do not require large
fractions of DM and they have been shown to describe well the dynamics 
of spiral and elliptical galaxies. 
By fitting the pressure profiles measured in the 
foreground clean SMICA map released by the Planck Collaboration,
we have found that clusters can also be accurately described
in these models. We have used a proprietary catalog of 579 clusters,
and have determined the parameter space that best fits data. Our results
are predicated on the baryonic
gas being in hydrostatic equilibrium in the potential wells
of clusters. This hypothesis can only be tested using
hydrodynamical simulations and if the gas turn out not to
be in equilibrium, our conclusion will be severely weaken. 
 
Models based on $f(R)$-gravity that do not require DM halos
appear as a viable alternative to generalized NFW models.
Due to foreground contamination, we cannot use 
single frequency maps. For instance, the 217GHz channel could
be used to remove the intrinsic CMB component and the signal 
at other frequencies could be fit to the profile of 
each individual clusters. Lacking frequency information
increases our error bars and makes our final contours wider
than what they would be otherwise. Then,
the constraints from pressure profiles could be further improve 
by using frequency information, by carrying out the analysis 
in foreground clean maps,
using the 217GHz map to remove the cosmological CMB signal
and fitting the profile of each individual cluster to the data.
The conclusion of this and similar studies \citep{cardone,napolitano}
is that large amounts of DM are not required to 
describe self-gravitating systems, if we relax the hypothesis 
that gravity is strictly scale independent above the
scale of Solar System.

\section*{Acknowledgments}

IDM and FAB acknowledge financial support from the Spanish
Ministerio de Educaci\'on y Ciencia (grant FIS2012-30926). SC and MDL 
acknowledge INFN (iniziative specifiche NA12 and OG51).

\end{document}